\documentclass[iop]{emulateapj}

\slugcomment{Accepted 22$^{\rm nd}$ June 2012}

\shorttitle{The Hungry Spider}
\shortauthors{Seymour et al.}

\begin{document}


\title{Rapid Coeval Black Hole and Host Galaxy Growth in MRC\,1138-262: The Hungry Spider}


\author{N. Seymour\altaffilmark{1,2}}
\altaffiltext{1}{CSIRO Astronomy \& Space Science, PO Box 76, Epping, NSW, 1710, Australia}
\altaffiltext{2}{Mullard Space Science Laboratory, University College London, Holmbury St Mary, Dorking, Surrey, RH5 6NT, UK}

\author{B. Altieri\altaffilmark{3}}
\altaffiltext{3}{Herschel Science Centre, European Space Astronomy Centre, ESA, Villanueva de al Ca\~nada, 28691 Madrid, Spain}

\author{C. De Breuck\altaffilmark{4}}
\altaffiltext{4}{European Southern Observatory, Karl Schwarzschild Stra\ss e 2, 85748 Garching, Germany}

\author{P. Barthel\altaffilmark{5}}
\altaffiltext{5}{Kapteyn Astronomical Institute, University of Groningen, PO Box 800, 9700 AV Groningen, The Netherlands}
\author{D. Coia\altaffilmark{3}}
\author{L. Conversi\altaffilmark{3}}

\author{H. Dannerbauer\altaffilmark{6}}
\altaffiltext{6}{Universit\"at Wien, Institut f\"ur Astronomie, T\"urkenschanzstra\ss e 17, 1180 Wien, \"Osterreich}

\author{A. Dey\altaffilmark{7}}
\altaffiltext{7}{NOAO-Tucson, 950 North Cherry Avenue, Tucson, AZ 85719, USA }

\author{M. Dickinson\altaffilmark{7}}
\author{G. Drouart\altaffilmark{8,4,1}}
\altaffiltext{8}{Institut d~Astrophysique de Paris, 98bis Bd Arago, 75014 Paris, France }
\author{A. Galametz\altaffilmark{9}}
\altaffiltext{9}{INAF - Osservatorio di Roma, Via Frascati 33, I-00040, Monteporzio, Italy}
\author{T.R. Greve\altaffilmark{10}}
\altaffiltext{10}{Department of Physics and Astronomy, University College London, Gower Street, London WC1E 6BT, UK}

\author{M. Haas\altaffilmark{11}}
\altaffiltext{11}{Astronomisches Institut, Ruhr-Universit\"at Bochum, Universit\"atsstr. 150, Geb\"aude NA 7/173, D-44780 Bochum, Germany}
\author{N. Hatch\altaffilmark{12}}
\altaffiltext{12}{School of Physics and Astronomy, University of Nottingham, University Park, Nottingham, NG7 2RD, UK}

\author{E. Ibar\altaffilmark{13}}
\altaffiltext{13}{UK Astronomy Technology Centre, Royal Observatory, Blackford Hill, Edinburgh, EH9 3HJ, UK}
\author{R. Ivison\altaffilmark{13,14}}
\altaffiltext{14}{Institute for Astronomy, University of Edinburgh, Royal Observatory, Edinburgh, EH9 3HJ, UK}
\author{M. Jarvis\altaffilmark{15,16}}
\altaffiltext{15}{Centre for Astrophysics Research, STRI, University of Hertfordshire, Hatfield, AL10 9AB, UK}
\altaffiltext{16}{Physics Department, University of the Western Cape, Bellville 7535, South Africa}

\author{A. Kov\'acs\altaffilmark{17}}
\altaffiltext{17}{University of Minnesota, 116 Church St SE, Minneapolis, MN 55414, USA}
\author{J. Kurk\altaffilmark{18}}
\altaffiltext{18}{Max Planck Institut f\"ur extraterrestrische Physik, Postfach 1312, 85741 Garching bei M\"unchen, Germany}

\author{M. Lehnert\altaffilmark{19}}
\altaffiltext{19}{GEPI, Observatoire de Paris, UMR 8111, CNRS, Universit\'e Paris Diderot, 5 place Jules Janssen, 92190, Meudon, France}
\author{G. Miley\altaffilmark{20}}
\altaffiltext{20}{Leiden Observatory, University of Leiden, P.O. Box 9513, 2300 RA Leiden, Netherlands}
\author{N. Nesvadba\altaffilmark{19}}
\author{J.I. Rawlings\altaffilmark{2}}
\author{A. Rettura\altaffilmark{21}}
\altaffiltext{21}{Jet Propulsion Laboratory, California Institute of Technology, Pasadena, CA 91109, USA}
\author{H. R\"ottgering\altaffilmark{20}}
\author{B. Rocca-Volmerange\altaffilmark{8}}
\author{M. S\'anchez-Portal\altaffilmark{3}}
\author{J.S. Santos\altaffilmark{3}}
\author{D. Stern\altaffilmark{21}}
\author{J. Stevens\altaffilmark{14}}
\author{I. Valtchanov\altaffilmark{3}}
\author{J. Vernet\altaffilmark{4}}
\author{D. Wylezalek\altaffilmark{4}}

\begin{abstract}
We present a detailed study of the infrared spectral energy distribution of 
the high-redshift radio galaxy MRC\,1138-26 at $z=2.156$, also known as 
\objectname[MRC 1138-26]{the Spiderweb Galaxy}.
By combining photometry from {\it Spitzer}, 
{\it Herschel} and LABOCA we fit the rest-frame $5-300\,\mu$m
emission using a two component, starburst and active galactic nucleus (AGN), 
model. The total infrared ($8-1000\,\mu$m) luminosity of this galaxy is 
($1.97\pm0.28$)$\times10^{13}\,L_\odot$ with ($1.17\pm0.27$) and 
($0.79\pm0.09$)$\times10^{13}\,L_\odot$
due to the AGN and starburst components respectively. 
The high derived AGN accretion rate of  $\sim20\,\%$ Eddington,
and the measured star formation rate (SFR) of $1390\pm150\,$M$_\odot$yr$^{-1}$,
suggest that this massive system is in a special phase of rapid central 
black hole {\em and} host galaxy growth, likely caused by a gas rich merger 
in a dense environment. 
The accretion rate is sufficient to power both the jets and the previously 
observed large outflow. 
The high SFR and strong outflow suggest this galaxy could potentially
exhaust its fuel for stellar growth in a few tens of Myr, although the likely 
merger of the radio galaxy with nearby satellites suggest bursts of star 
formation may recur again on time scales of several hundreds of Myr.
The age of the radio lobes implies the jet started
after the current burst of star formation, and therefore we are possibly
witnessing the 
transition from a merger-induced starburst phase to a radio-loud AGN phase.
We also note tentative evidence for [CII]$158\mu$m emission. 
This paper marks the first results from the {\it Herschel} Galaxy Evolution 
Project (Project HeRG\'E), a systematic study of the evolutionary state 
of 71 high redshift, $1<z<5.2$, radio galaxies.

\end{abstract}


\section{Introduction}

There is now a growing consensus that high redshift radio galaxies (HzRG; 
$z>1$, $L_{\rm 500MHz}>10^{26}\,$WHz$^{-1}$) are markers of the early 
formation of the most massive galaxies, located within the highest peaks of 
dark matter over-densities \citep{Miley:08}. Not only do HzRGs tend to 
reside in significantly over-dense environments 
\citep[e.g.,][]{Ivison:00,Stern:03,Venemans:07,Galametz:10b,Tanaka:11,Galametz:12,Mayo:12}, 
but the mass of their  stellar component also tends to be high as indicated by 
the remarkably tight scatter between the observed $K-$band magnitude 
and redshift \citep[e.g.,][]{Lilly:84,Eales:97,Jarvis:01,Willott:03,RoccaVolmerange:04,Bryant:09b}. 
The high stellar masses have been confirmed more directly with rest-frame 
near-infrared observations \citep{Seymour:07,DeBreuck:10}. 
Recent observations of the high mid-infrared 
luminosities of HzRGs indicate that the black holes in these 
sources likely have high accretion rates 
\citep{Ogle:06,Seymour:07,Fernandes:11}.
HzRGs are often found with evidence of high star formation rates (SFRs) 
from detections in the sub-millimetre 
\citep[e.g.,][]{Archibald:01,Stevens:03,Reuland:04}, 
from their extended Lyman\,$\alpha$ emission indicating large reservoirs of 
ionized gas
\citep[e.g.,][]{Reuland:03,VillarMartin:03}, from their 
CO emission indicating reservoirs of molecular $H_2$ 
\citep[e.g.,][]{Papadopoulos:00,Emonts:11}, 
from hot star ultraviolet 
absorption lines \citep[e.g.,][]{Dey:97}
and from their mid-infrared spectra \citep[e.g.,][]{Seymour:08b,Ogle:12}.
Kinematic measurements from ionized gas have also shown that HzRGs have 
extremely massive central black holes 
\citep[$3\times10^{9}-2\times10^{10}\,M_\odot$ --][]{Nesvadba:11} 
consistent with the observed high stellar masses and accretion rates. 
All of these observations confirm the picture that HzRGs are markers of
massive galaxies co-forming with a central black hole, 
residing in peaks of dark matter over-density.

The exact formation mechanism of massive galaxies, and how the frequently 
observed
radio-loud phase is related to black hole and galaxy growth, is often debated.
A connection between the growth of the central black hole and the host galaxy
has been proposed to reconcile semi-analytical models with observations. 
A form of `feedback', associated with a radio-loud phase, has been proposed 
to truncate the growth of the bulge and central black hole growth,
thereby preventing the over-production of massive galaxies being formed in 
these simulations \citep{Bower:06,Croton:06}. This idea has gained popularity 
with the kinematic observations suggesting powerful outflows from many HzRGs
\citep[e.g.,][]{Nesvadba:06}, although most models favor scenarios where the 
winds are radiatively driven.

In order to test how phases of powerful radio-loud emission influence, or
are associated with, the 
formation of massive galaxies at high redshift, accurate measurements of the
black hole accretion rate and the SFR are required. Due to varying, and 
difficult to determine, levels of obscuration at UV and optical wavelengths, 
as well as contamination from the active galactic nucleus (AGN), the 
best wavelength regime in which to obtain such measurements is the infrared 
(IR) where the spectral energy distributions (SEDs) of such sources also 
tend to peak. With the {\it Spitzer Space Telescope} \citep{Werner:04} we 
conducted a systematic study with eight photometric bands ($3.6-160\,\mu$m) 
of 71 HzRGs with redshifts, $z$, in the range $1<z<5.2$ in order to measure 
their near- to mid-IR rest-frame SEDs. This photometry allowed us to estimate 
separately the stellar and AGN contribution at these wavelengths, thereby 
confirming their high stellar masses and AGN mid-IR luminosities 
\citep{Seymour:07,DeBreuck:10}. In order to measure the complete IR SED, 
we are now extending this study to far-IR wavelengths as the {\it Herschel}
Radio Galaxy Evolution Project ({\it Projet HeRG\'E}). The {\it Herschel Space
Telescope}\footnote{{\it Herschel} is an ESA space observatory with 
science instruments provided by European-led Principal Investigator consortia 
and with important participation from NASA.}  
\citep{Pilbratt:10} has already demonstrated that far-IR 
measurements can reliably be used to measure SFRs in powerful AGN 
\citep[e.g.,][]{Hardcastle:10,Hatziminaoglou:10,Shao:10,Seymour:11}.
By combining these data with those from {\it Spitzer} we can, for the first
time, accurately measure the total bolometric luminosities and far-IR
contribution, and then estimate the relative AGN and starburst IR luminosities 
in HzRGs. 

In this {\it letter} we present the first results from {\it Projet HeRG\'E} 
on MRC\,1138-262 at $z=2.156$, otherwise known as the
Spiderweb Galaxy 
due to the preponderance of satellite 
galaxies, `flies', revealed by deep {\it Hubble Space Telescope (HST)} 
imaging \citep{Miley:06}. This HzRG has a powerful AGN as implied by its 
high X-ray luminosity 
\citep[$L_{\rm 2-10keV}=4\times10^{45}\,$erg\,s$^{-1}\sim1.0\times10^{12}\,L_\odot$ --][]{Carilli:02a}, 
its extreme radio emission 
\citep[$L_{\rm 500MHz}=1.2\times10^{29}\,$WHz$^{-1}$ --][]{DeBreuck:10},  
and evidence of a high SFR 
\citep[$\sim1200\,M_\odot$yr$^{-1}$ --][]{Stevens:03,Ogle:12}. 
There is also evidence of a powerful outflow \citep{Nesvadba:06} potentially 
driven by the massive central black hole \citep{Nesvadba:11}.
A study of the proto-cluster environment of the Spiderweb Galaxy shows that 
many members will be stripped
by, or merge with, the central galaxy by the present epoch 
\citep{Hatch:09,Doherty:10,Tanaka:10}. \citet{Kuiper:11} suggest that the 
measured dynamical structure around the Spiderweb Galaxy can be interpreted 
as the merger of two massive haloes and that the combined system has not yet 
become viralised.
In this paper we aim to measure more accurately the AGN accretion rate and 
SFR, the bolometrically dominant components, thereby obtaining an improved 
understanding of the evolutionary state of this HzRG.

In \S\ref{data} we present the IR data used in this work and in \S\ref{fit} we
present the SED fitting of these data with an AGN and starburst component. 
We discuss our results, and the overall evolutionary state of the Spiderweb
Galaxy in \S\ref{discussion} and conclude the paper in \S\ref{conclusion}. 
Throughout this paper we use the
`concordance' $\Lambda$CDM cosmology with $H_0=70\,$Mpc$^{-1}$kms$^{-1}$, 
$\Omega_\Lambda=1-\Omega_M=0.7$.

\section{Data}
\label{data}

The data used in this work comprises observed infrared photometry across
$20-1000\,\mu$m. At the shortest wavelength, $24\,\mu$m, the photometry
comes from {\it Spitzer} and was published in \citet{DeBreuck:10}.
We add additional
contributions to the reported uncertainty, in quadrature, to allow for
the absolute flux calibration and the colour correction uncertainty 
as described in \citet{Seymour:07}.
We supplement this photometry with new observations from {\it Herschel} 
using the PACS \citep{Poglitsch:10} and SPIRE \citep{Griffin:10} 
instruments, as well as a LABOCA/$870\,\mu$m flux density
(Dannerbauer et al., in prep., ESO program ID 084.A-1016(A) and MPG 
program ID 083.F-0022 -- both PI J. Kurk).
The {\it Herschel} observations were carried out 
as part of the project scientist guaranteed time (PI Altieri).

The SPIRE observation (OBSID 1342210877) was performed using a four 
repetition cross-scan Large Map mode covering $10\times 10\,$arcmin, 
achieving $1\,\sigma$ instrumental noise of (3.5, 4.8, 5.4) mJy at (250, 
250, 500$\,\mu$m) --- slightly below the SPIRE extragalactic confusion 
noise \citep{Nguyen:10}. To obtain SPIRE photometry for the Spiderweb 
Galaxy we used a time-line fitter\footnote{The time-line fitter for SPIRE is 
available at the {\it Herschel}/SPIRE public wiki page: 
\url{herschel.esac.esa.int/twiki/bin/view/Public/SpireCalibrationWeb}}, 
which provides a better estimate of the flux density, because no binning 
into sky pixels is involved in the process. The final flux density 
uncertainty, reported in Table~\ref{tab:data}, includes the 
fit error, the confusion noise and a conservative $7\,\%$ flux calibration 
uncertainty (see the SPIRE Observers' Manual).
The photometry from standard source fitting on maps is consistent with the 
time-line fitter results, but we prefer to use the time-line fitter results 
as these are generally more accurate, especially for sources at 
high signal-to-noise.
The PACS observations (OBSID 1342222456 and 1342222457) 
consisted of two crossed scan maps covering the same area
and the data were reduced using HIPE v6.0 \citep{Ott:10}. The 
data cubes were processed with a standard pipeline, including a sliding 
high-pass filtering on the detector time-lines to remove detector drifts 
and 1/f noise with an iterative masking of the sources. 
We extracted flux densities via standard aperture photometry techniques and
the PACS uncertainties 
include contributions from instrumental noise and absolute flux calibration.
For the LABOCA/$870\,\mu$m flux density uncertainty we add a $15\,\%$ 
uncertainty, in quadrature, to the instrumental uncertainty to account 
for absolute flux calibration, atmospheric effects.

\begin{figure*}[h]
   \centerline{
   \includegraphics[angle=0,scale=.50]{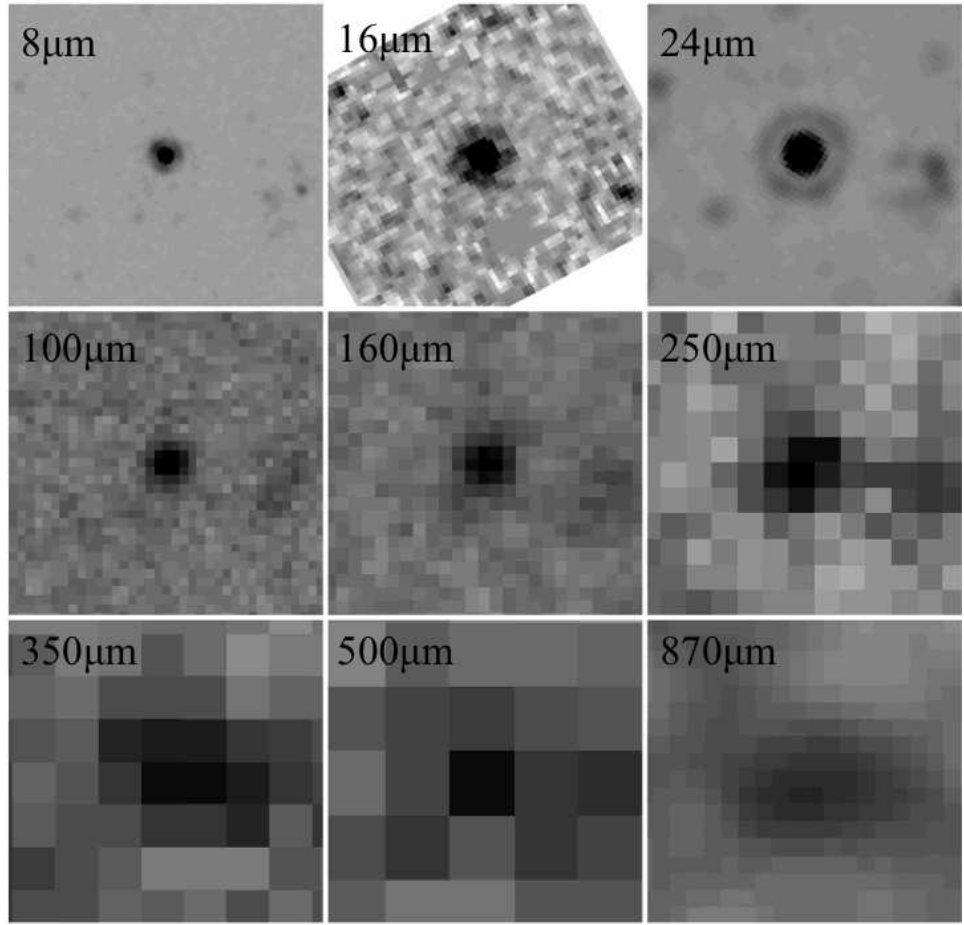}
   }	
   \caption{Postage-stamp IR images, $70\times 70\,$arc-seconds, of the 
     Spiderweb Galaxy (RA 11:40:48.60, Dec. $-26$:29:8.50; J2000) 
     with North up and East to the left. From top left to bottom right 
     the images are: {\it Spitzer}/$8/16/24\,\mu$m, PACS/$100$ and 
     $160\,\mu$m, SPIRE $250/350/500\,\mu$m and LABOCA/$870\,\mu$m. 
     The $8$ and $16\,\mu$m band are shown for illustration only and are 
     not used in the model fitting. Note the Airy ring 
     is quite prominent in the {\it Spitzer} data.}
\label{fig:postage}
\end{figure*}

The {\it Herschel} beam varies from $6-36\,$arc-seconds, but the radio 
galaxy appears point like in all-bands from $8\,\mu$m to $250\,\mu$m
(see Fig.~\ref{fig:postage}). The 350, 500 and $870\,\mu$m images all show a 
slight extension in the east-west direction consistent with emission from the 
clump of galaxies approximately 30\,arc-seconds to the west of the radio 
galaxy seen in the $8\,\mu$m through $250\,\mu$m bands. This extension is
consistent with the SCUBA/$850\,\mu$m detection  by \citet{Stevens:03}. 
This clump comprises of four sources seen individually at $8\,\mu$m
and $24\,\mu$m, the brightest two of which have been spectroscopically 
confirmed to lie at the same redshift and therefore be proto-cluster members 
\citep{Kurk:04a}; hence, we can assume the long-wavelength flux density from 
these sources to have similar colours due to the Rayleigh-Jeans tail. We note 
the peak flux density at the longer wavelengths remains centered on the radio
galaxy. We assume, for the purposes of SED fitting, that all this flux density 
comes from one source, but then treat the SFR measured from the starburst 
component as an upper limit to that from just the radio galaxy.

All the data used in the fitting of this sample are presented in 
Table~\ref{tab:data}. 

\section{SED fitting}
\label{fit}

The goal of our SED fitting is to obtain the AGN and starburst IR luminosities
($8-1000\,\mu$m) and then derive physical quantities from these results.
We fit the observed photometry across $5-300\,\mu$m rest frame with a 
two-component model consisting of an AGN and a starburst template. 
Extrapolation of the radio core flux density using the observed flux densities 
at wavelengths $\ge6\,$cm and the core spectral index from \citet{Carilli:97}, 
$\alpha=-1.3$ ($S_{\rm \nu}\propto \nu^\alpha$), suggests that synchrotron 
emission at these wavelengths is many orders of magnitude fainter.
We use the {\tt Decompir}\footnote{\url{https://sites.google.com/site/decompir/}} 
SED model fitting code \citep{Mullaney:11} to fit the two components
with some minor adaptations to use additional band-pass filters. 
The AGN component is an empirical model based on observations of 
moderate-luminosity (i.e., $L_{\rm 2-10keV}=10^{42}-10^{44}\,$erg\,s$^{-1}$) 
low-redshift 
AGN in \citet{Mullaney:11} selected at high X-ray energies from 
the local, $z<0.1$, {\it Swift}-BAT sample \citep{Tueller:08}. 
This model is a continuous function comprised of two power-laws
and a modified black-body, expressed as:

\begin{equation}
F_\nu \propto \left\{
\begin{array}{rcl}
  \lambda^{1.8} & \mbox{at} & 6\,\mu{\rm m} < \lambda < 19\,\mu{\rm m}\\
  \lambda^{0.2} & \mbox{at} & 19\,\mu{\rm m} < \lambda < 40\,\mu{\rm m} \\
  \nu^{1.5}F_\nu^{\rm BB} & \mbox{at} & \lambda > 40\,\mu{\rm m} \\
\end{array}\right.
\label{Eqn:AGN}
\end{equation}

\noindent
In comparison with other AGN templates commonly used, it is broadly similar 
in the mid-IR regime; however it has lower far-IR 
emission, when normalized at $19\,\mu$m, compared to NGC 1068 and Mrk 231,
likely attributable to unaccounted starburst 
contributions at far-IR wavelengths in these two galaxies \citep{LeFloch:01}. 
This model is also broadly similar to empirical models presented by 
\citet{Richards:06} and \citet{Netzer:07}.
The advantage of the formulation above is that the various parameters (the 
two power-law slopes and the black-body peak wavelength) can be adjusted to 
closely match these other models which we test later on. 

The starburst models consist of five SEDs developed by Mullaney et al. from 
\citet{Brandl:03} designed to represent a typical range of SED types (i.e., 
peaking at a range of wavelengths). Beyond $100\,\mu$m these templates
are extrapolated using a grey body with a frequency dependent emissivity 
approximated as $\sim\nu^\beta$ with $\beta=1.5$. In addition to the 
normalisation and choice of starburst template, {\tt Decompir} can apply 
extinction to each component separately. As the AGN-dominated mid-IR spectrum 
\citep{Ogle:12} shows a little silicate emission, not absorption, we only 
allow the starburst component to be affected by absorption\footnote{We note 
that while the X-ray emission from AGN shows moderate extinction, 
$N_H=$($2.6\pm0.6$)$\times10^{22}$cm$^{-2}$ 
\citep{Carilli:02a}, the mid-IR SED does not \citep{Ogle:12}. 
This is consistent with the bright H\,$\alpha$ seen by \citet{Nesvadba:06}
and can be attributed to differences in dust grains seen toward the 
circumnuclear regions of AGN and the diffuse ISM of our galaxy.}
Therefore the chi-squared, $\chi^2$, 
minimisation performed by {\tt Decompir} has four free components: 
normalisation of each template, choice of starburst template and starburst 
extinction. 

\begin{figure*}[h]
   \centerline{
   \includegraphics[angle=-90,scale=.50]{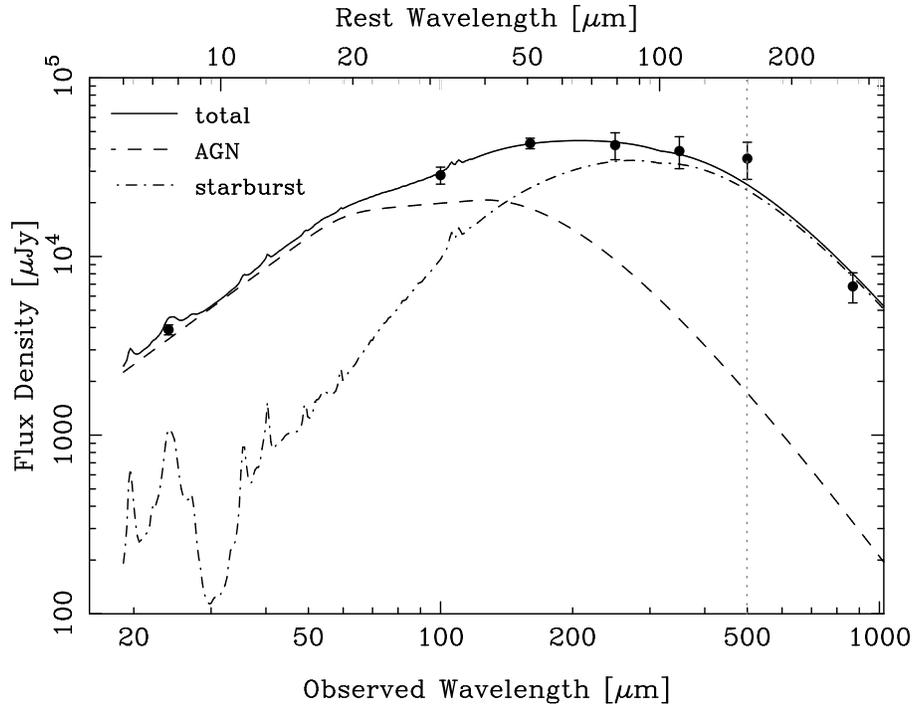}
   }	
   \caption{SED fit to the rest-frame $5-300\,\mu$m rest-frame photometry 
     (black dots) of 
     the Spiderweb Galaxy at $z=2.156$ (MRC1138-262). The total model fit, and 
     AGN and starburst components are indicated in the figure. 
     The vertical grey dotted line represents the 
     position of the [CII]158$\mu$m line. }
\label{sed}
\end{figure*}

The best fit combination of models returns a reduced chi-squared of 
$\chi_r^2=1.4$ 
with the starburst template, `SB4', the template peaking at the shortest 
wavelength. We note that while the best fit model has some extinction
for the starburst component, it is poorly constrained and we get an 
equally good fit with no extinction and similar luminosities within $2\,\%$.
The fit is shown in Fig.~\ref{sed} and the luminosities are reported in 
Table~\ref{tab:res}. The model fits the 
data well, although at the longest wavelengths the model slightly struggles
to fit the sharp decrease in flux density from $500$ to $870\,\mu$m. The SED
at these wavelengths is dominated by the Rayleigh-Jeans tail of the starburst 
component and any blackbody would struggle to fit such a flux density ratio, 
although the slope does depend on the emissivity index.
The `SB4' starburst template is the hottest of the five 
used, with a dust temperature of $T_{\rm dust}\sim 64\,$K. 
The next two cooler templates, $T_{\rm dust}\sim 46-53\,$K,
give a $\chi_r^2\approx 5-8$ and fit the $500\,\mu$m flux density better, but 
exceed the $870\,\mu$m flux density by $\ge4\,\sigma$. 

If we exclude the $500\,\mu$m data point we marginally improve the best 
fit to $\chi_r^2=1.1$, again for the `SB4' template and with the output 
parameters changing very little (note the number of degrees of freedom drops 
by one). One potential method to better fit the observed high $500$ to 
$870\,\mu$m flux ratio is to use starburst SEDs with a steeper emissivity 
index. By modifying the templates so that 
$\beta=2$ we can obtain a steeper slope on the Rayleigh-Jeans side of the 
SED peak. Repeating the SED fitting with these modified starburst templates
we obtain a best fit of $\chi_r^2=1.1$ from the modified `SB5' starburst 
template. The overall fit does not significantly change with a slightly 
smaller $500\,\mu$m excess, by $\sim20\,\%$,
as well as an AGN luminosity increase of $\sim10\,\%$ and a 
starburst luminosity decrease of $\sim20\,\%$. Hence, there is no significant
improvement with this change to the starburst templates.

Visual examination of the images at 350, 500 and $870\,\mu$m, see 
Fig.~\ref{fig:postage}, suggests the measured flux densities at these 
wavelengths may have contributions from other proto-cluster members as 
discussed in \S\ref{data}. As the galaxies responsible for this emission 
lie at the same redshift, and are almost certainly starbursts to be 
emitting at these wavelengths, they likely have similar flux density ratios.
The $250\,\mu$m flux density 
is not affected by these nearby sources and the peak flux density still 
originates from the central radio galaxy. The ratio of the radio galaxy to 
companion peak flux densities at $250\,\mu$m is about $2:1$; hence, we 
estimate the contribution of the companion to the measured flux density
at longer wavelengths to be $\sim\frac{1}{3}$ at most. We note that the 
$350\,\mu$m flux density need only be decreased by $25\,\%$ to produce an 
unphysical flux density ratio for a cold dust SED. However,  
if we decrease the three longest wavelength flux densities by this much 
the best fit gives a starburst luminosity only $10\,\%$ lower. Hence, 
although the estimated starburst IR luminosity is formally an upper limit
for the radio galaxy, we postulate that the true starburst IR luminosity 
of the Spiderweb Galaxy is, at most, only $10\,\%$ lower. 

We do not have enough photometry at 
$5\,\mu$m$\,\le\lambda_{\rm rest}\le50\,\mu$m 
to allow all the AGN parameters (the two power-law slopes and the black-body 
peak), other than the normalisation, to vary simultaneously. 
Allowing just one parameter to vary results in changes of luminosities of 
$<5\,\%$. We can change the parameters in equation one to mimic the 
\citet{Richards:06} and \citet{Netzer:07} AGN models. In both cases, the
AGN luminosity halves and the starburst luminosity doubles, but the fit is 
worse with $\chi_r^2\sim8.5$. We note that this fit is improved if we use 
the Richards or Netzer AGN model with the $\beta=2$ starburst SED, with the
same qualitative results, but with a $\chi_r^2=3.3$.
We also tried using the suite of empirical starburst templates from 
\citet{Rieke:09} instead of the Brandl templates. The best fit templates
were also those peaking at the shortest wavelengths, i.e., the ultra
luminous IR galaxy (ULIRG) templates,
but with a slightly higher chi-squared: $\chi_r^2\sim3$.
To obtain a better understanding of the uncertainties we also randomly varied 
the flux densities with a Gaussian distribution based on their quoted 
uncertainties one hundred times and repeated the fitting procedures. We found 
the variance of the resulting luminosities to be $\le10\,\%$, i.e., less than 
the quoted uncertainties in Table~\ref{tab:res}.

We finish this section by noting that while there are many ways to tweak 
both the AGN and starburst templates and change the parameters of the fit all 
solutions have an AGN component dominating at the shortest wavelengths, 
$\lambda_{\rm rest}\le30\,\mu$m, and a starburst component dominating at 
long wavelengths $\lambda_{\rm rest}\ge60\,\mu$m. Hence, all good fits have a
powerful, $L_{\rm IR}$(SB/AGN)$>5\times10^{12}\,L_\odot$, starburst and 
AGN component. Until we get high spatial and/or spectral resolution data, 
e.g. from ALMA or JWST, it will be difficult to differentiate between many 
moderately good fits.

\section{Discussion}
\label{discussion}

We have measured very high AGN and starburst luminosities which imply a high 
central black hole accretion rate and SFR, respectively. The total IR 
luminosity of this system causes it to be classed as a hyper-luminous IR 
galaxy, $L_{\rm IR}>10^{13}\,L_\odot$ . We discuss here what these 
results mean and how they relate to the radio emission and overall evolutionary
state of the system. We also discuss the potential origin of the $500\,\mu$m
excess.

\subsection{AGN Power}

The total IR AGN luminosity marks the Spiderweb Galaxy 
as one of the most powerful AGN known; therefore the AGN is likely having 
a strong effect on its host galaxy and immediate environment 
(see \S\ref{dis:env}). 
Assuming the full unobscured AGN SED is similar to the \citet{Elvis:94} 
templates, we get $L_{\rm BOL}$(AGN) $\sim6\times\,L_{\rm IR}$(AGN) (we note
other unobscured AGN templates produce similar numbers) and 
therefore an AGN bolometric luminosity of $\sim7\times10^{13}\,L_\odot$. 
This bolometric AGN luminosity estimate is higher than previous estimates, 
e.g., \citet{Nesvadba:06} who used the 
observed X-ray luminosity \citep{Carilli:02a} and corrected it using the 
\citet{Elvis:94} template, $L_{\rm BOL}$(AGN) 
$\sim10\times\,L_{\rm 2-10keV}$(AGN). 

This discrepancy is unlikely due to extinction as seen towards the nucleus 
in X-rays given the low absorbing column \citep{Carilli:02a}.  As noted 
earlier, neither the mid-IR spectrum \citep{Ogle:12} nor the SED 
(Drouart et al. 2012) indicate significant extinction either.  The 
discrepancy between the hard X-ray and IR-derived bolometric AGN 
luminosities is more likely due to the dispersion between individual 
sources and the \citet{Elvis:94} templates.  This, in turn, could be 
explained by a different dust-to-gas ratio in this source, a clumpy 
torus or an unusual AGN geometry, but further investigation is beyond the 
scope of this paper. 

This galaxy exhibits both a large scale central velocity distribution 
indicative of a powerful outflow \citep[][]{Nesvadba:06} and a 
high-luminosity mid-IR $H_2$ molecular emission line \citep{Ogle:12}. 
The derived bolometric radiative power would therefore be enough to power 
both the outflow \citep[$\sim8\times10^{12}L_\odot$ --][]{Nesvadba:06} and the 
probably shock-heated gas \citep[$\sim4\times10^{10}L_\odot$ --][]{Ogle:12}, 
even assuming a $10\,\%$ coupling efficiency. However, the kinetic 
power of the radio jet, estimated to be $0.6-8.0\times10^{13}\,L_\odot$ by 
\citet{Nesvadba:11}, is also powerful enough to potentially drive the outflow. 

Nesvadba et al. (2006) also present an estimate of the black hole mass from 
the measurement of the central H\,$\alpha$ luminosity and FWHM.
The principle uncertainty in this measurement is how the orientation 
of the galaxy affects the application 
of the \citet{Greene:05} scaling relationship, which is based on a sample of 
lower luminosity AGN likely to be more face-on than the Spiderweb galaxy, 
i.e., at lower inclinations.
Indeed, G. Drouart et al. (2012, in preparation) find a correlation 
between radio core prominence and viewing angle from a hot torus model fitted
to the mid-IR SED of HzRGs which suggests that the radio axis is aligned to 
torus axis. These authors find that the Spiderweb Galaxy 
is inclined at angle of $\sim68\deg$ to the line of sight, but has modest
extinction, an $A_V\sim1.5$. While generally there may be misalignments of 
up to $10-20\deg$ between the radio and torus axes, it is likely that the 
Spiderweb Galaxy is more edge-on than the \citet{Greene:05} sample.
\citet{McLure:02a} show that the measured FWHM from a thin accretion disk is 
proportional to the sine of the inclination angle squared. Therefore, if the 
Spiderweb Galaxy has a higher inclination angle, $\theta_{\rm RG}$, than the 
mean value for the AGN sample used in \citet{Greene:05}, $\theta_{\rm GH05}$, 
then we would be overestimating the black hole 
mass by $\frac{\sin^2(\theta_{\rm GH05})}{\sin^2(\theta_{\rm RG})}$ which 
\citet{Nesvadba:11} assume to be $\sim2$. We therefore take the 
lower, inclination-corrected, value of the black hole mass from 
\citet{Nesvadba:11}, $\sim10^{10}\,M_\odot$.
The higher bolometric luminosity derived above implies a higher 
black hole mass derived from the H\,$\alpha$ luminosity.
However, we also note the black hole mass reported here would
put this galaxy on par with some of the most massive black holes known 
in the local Universe \citep{McConnell:11}, and are very rare for the $z>1$
Universe. 
Hence, we treat this black hole mass as uncertain by a factor $2-3$.

We then relate the bolometric luminosity to the maximally possible accretion 
luminosity where the continuum radiation force outwards balances the 
gravitational force inwards in hydrostatic equilibrium,
i.e., the Eddington luminosity, and we find an Eddington 
accretion rate of $\sim20\,\%$. This is a factor of a few
higher than the estimate of \citet{Nesvadba:11} due to the 
higher bolometric 
luminosity and lower black hole mass used in this work. 
This high rate of accretion is common in less bolometrically luminous AGN,
\citep[$\sim25\,\%$ --][]{Kollmeier:06}, 
but is rare for less radio-luminous AGN 
\citep[$L_{\rm 500MHz}\lessapprox10^{27}\,$WHz$^{-1}$ --][]{Best:12} at 
least in the local, $z<0.1$, Universe.
However, it is probably common for HzRGs as based on their X-ray 
\citep{Overzier:05} and mid-IR \citep{Ogle:06,Seymour:07} luminosities. 
These results suggest the accretion disk is in an efficient state, 
although the large outflow hints that it may soon run out of fuel. 

\subsection{Star-Formation Rate}

The starburst IR luminosity implies a SFR of $1390\pm150\,$M$_\odot$yr$^{-1}$ 
using the \citet{Kennicutt:98} scaling relation (and therefore a Salpeter 
IMF). We noted in \S\ref{data} that the longer wavelength bands 
may have some contribution from nearby proto-cluster members, but 
estimate that this contribution is $\le10\,\%$ in \S\ref{fit}.
This SFR is much higher than estimates from 
optical/near-IR data \citep{Hatch:08,Hatch:09}, likely due to 
obscuration by dust, 
but is broadly consistent with estimates from 
the PAH luminosities measured from the mid-IR SED 
(Ogle et al., 2012; J. Rawlings, 2012, in preparation).
We note that the IR luminosity is more sensitive to the star formation 
averaged over a slightly longer time frame than optical/UV 
observations. We can estimate the mechanical energy of this starburst using 
the canonical conversion factor from \citet{DallaVecchia:08} and obtain a 
value $1-2\times10^{11}\,L_\odot$ which 
is insufficient to explain the kinematics and outflow observed 
\citep{Nesvadba:06}. However, any material driven out may fall 
back onto the central galaxy or the satellite galaxies over time
\citep[e.g.,][and see \S\ref{dis:env}]{Hatch:09}.

Estimating the stellar mass 
of this source is difficult due to the strong AGN continuum. \citet{Hatch:09} 
provide an estimate of $1.1\pm0.2\times10^{12}\,M_\odot$ using SED fitting 
of high resolution {\it HST} rest-frame UV/optical data. This result is 
consistent with the upper limit found from rest-frame near-IR observations 
\citep{Seymour:07,DeBreuck:10}. We adopt the \citet{Hatch:09} value as an 
upper limit since it may still be contaminated by the AGN as it lies in 
the AGN-dominated region of mid-IR colour-colour space as shown in 
\citet{DeBreuck:10}. We can now 
derive the specific SFR, sSFR=SFR/$M_{\rm stellar}$, and find a value of 
$\gtrsim1.3\,$Gyr$^{-1}$. This is a high sSFR, but common for
star forming galaxies at this redshift \citep{Noeske:07b,Elbaz:11}. However, 
it is unusual for a galaxy this massive. 

The inverse of the sSFR gives the characteristic time, $t_{\rm char}$, 
in which the galaxy would double its stellar mass if it maintains 
the current SFR. While this time-frame is short, $\lesssim0.8\,$Gyr, for the 
Spiderweb, it is likely that the star formation 
will turn off on even shorter timescales. Measurements of CO 
provide the best estimates of the mass of molecular hydrogen, i.e., 
the main component of the fuel for star formation and black hole growth 
\citep[e.g.,][]{Emonts:11}. 
While such measurements do not currently exist for the Spiderweb Galaxy 
we infer a value of $M_{\rm H_2}\sim0.5-5\times10^{10}\,M_\odot$ 
based the far-IR luminosity and the conversion factors used in 
\citet{Emonts:11} assuming the Spiderweb is similar to the HzRGs
in that study.
If we assume $100\,\%$ star formation efficiency, then this SFR can only be 
maintained for another $4-40\,$Myr, neglecting infall from satellite 
galaxies which may happen within a few hundred Myr (see below). 
If the mass of material estimated to being removed by the outflow, 
$\gtrsim400\,M_\odot$yr$^{-1}$ \citep{Nesvadba:06}, is taken into account, this
time frame decreases by $\sim30\,\%$, more if the powerful jets also entrain 
and remove significant amounts of gas from the galaxy, or if the molecular
hydrogen mass is lower than assumed. 
If we assume a much lower star formation efficiency, e.g. $10\,\%$, then 
this time-scale decreases by an order of magnitude, but leaves fuel for
further bursts of star-formation triggered by its proto-cluster environment.

\subsection{Age of Radio Jet}

Previously, \citet{Nesvadba:06} estimated the age of the radio jet 
using the method put forth by \citet{Wan:00}
and obtained a value of $4-10\,$Myr. We can 
re-estimate the age using models of radio lobe evolution from \citet{Mocz:11a}.
These authors reproduce the emission at radio wavelengths from synchrotron 
processes and the X-rays emitted by the lobes due to inverse Compton (IC) 
scattering
of the Cosmic Microwave Background. These models lead to tracks of lobe 
radio/$151\,$MHz and X-ray/$1\,$keV luminosity against time and size of radio
lobes at different redshifts. 
\citet{Carilli:02a} estimate that $10-25\,\%$ of the X-ray emission 
they measure is spatially extended on scales of $10-20\,$arcsec, similar to the
extent of the radio jet, $15.8\,$arcsec \citep{Carilli:97}. Hence, if we
assume all this X-ray luminosity is from IC processes we get a lobe
X-ray luminosity of $L_{\rm 2-10keV}\sim1-2.5\times10^{11}\,L_\odot$. As 
\citet{Mocz:11a} use a 1\,keV X-ray luminosity we take 
$L_{\rm 1keV}\sim1.8-4.5\times10^{11}\,L_\odot$ (assuming a photon power-law 
index of $\Gamma=1.4$), and from the 365 and 74\,MHz radio flux densities 
\citep{Douglas:96,Cohen:07} we find a $151\,$MHz luminosity of 
$L_{\rm 151}=5.6\times10^{29}\,$WHz$^{-1}\sim2.2\times10^{11}\,L_\odot$. 
Using the redshift and inclination estimate from G. Drouart et al. (2012, 
in preparation) we estimate the total lobe size to be $\sim140\,$kpc, which
implies an age of $\sim5\,$Myr in Fig.~1 of \citet{Mocz:11a} using the 
more powerful, $z=2$ AGN track. Using the radio and X-ray luminosities, 
where $L_{\rm 151}\ge L_{\rm 1keV}$, in Fig.~3 of Mocz et al.,
we obtain an estimate of $\le 25\,$Myr for the age
of the lobe. Hence, we take the age of the lobes to be in the range
$5-25\,$Myr.

\subsection{Possible [CII]158$\mu$m Line}
\label{dis:cii}

In \S\ref{fit} we reported an excess of emission in the $500\,\mu$m band 
which we measure to be 
$10.0\,$mJy at $\ge1\,\sigma$ significance ($\sim2\,\sigma$ based on just the 
instrumental noise). A potential explanation for this slight excess 
could be a significant contribution from the redshifted [CII]158$\mu$m line 
($\lambda_{\rm obs}=499\,\mu$m). We note that no other lines, e.g. CO, fall 
within the $500\,\mu$m 
band width. \citet{Smail:11} examine the effect of strong far-IR 
emission lines on broad band photometry and show that such emission
lines can cause measurable excesses in the broad band photometry. The excess 
measured here, $40\%$ above the $25.3\,$mJy of the model, 
corresponds to a value of $L_{\rm [CII]}/L_{\rm IR}\sim2.7\,\%$ 
\citep[e.g. Fig. 4 of ][]{Smail:11}.
While this value is higher, by a factor of $\sim2$, than any known value, 
it potentially indicates evidence for strong [CII]158$\mu$m 
distributed across the proto-cluster environment, due to the extended
$350-870\,\mu$m emission in the direction of confirmed proto-cluster members.
The idea of a high, 
extended [CII] luminosity is consistent with the idea of a gaseous 
environment of neutral gas, ionized gas \citep{Kurk:03}, shocked $H_2$ 
\citep{Ogle:12} and intra-galaxy star formation 
\citep{Hatch:09} within this proto-cluster environment. 
We caution that the statistical significance of this result is poor 
but the Spiderweb Galaxy remains a promising target for 
far-IR/millimetre spectroscopy.

\subsection{Evolutionary State of the Spiderweb Galaxy}

We have determined that the Spiderweb Galaxy is in a rare, short-lived, 
phase of extreme, concurrent black hole and host galaxy growth. If we compare 
the relative luminosities of the AGN and starburst component to the 
`continuous growth' relation in \citet{Lutz:10}, i.e., the ratio required to 
maintain a constant black hole to stellar mass ratio, we see that the black 
hole is growing at a relatively faster rate by almost 1\,dex. This result is 
consistent with the observed, albeit rather uncertain, ratio of stellar to 
black hole mass, $\le110$, i.e. the black hole is relatively more
massive (and has been growing rapidly) than it would be for the Spiderweb
Galaxy to lie on the local `$M-\sigma$' mass ratio of $\sim200$ 
\citep{Magorrian:98}. The stellar and black hole  masses do lie on the 
relationship found by \citet{Haring:04} and we note that both observed 
relations have a large scatter, $0.3\,$dex. 
If this source is to end up on the local `$M-\sigma$' relation, further 
build-up of stellar mass is required. However, we have estimated that 
the star formation rate will likely rapidly decrease on 
time-scales of $\sim40\,$Myr assuming no further source of fuel.

\citet{Page:12} have recently examined the {\it Herschel}/$250\,\mu$m
detection rate of X-ray luminous AGN in the {\it Chandra} Deep Field North
in the $1<z<3$ redshift range as a function of X-ray luminosity. Those authors 
found that the fraction of X-ray AGN detected with 
$S_{\rm 250\,\mu m}\ge18\,$mJy decreases to zero above 
$L_{\rm 2-8keV}=10^{44}\,$erg\,s$^{-1}$ ($=3\times10^{10}\,L_\odot$) and 
interpreted this result as suppression of star formation by powerful outflows
driven by the AGN. The Spiderweb is a clear counter-example of this result
with an X-ray luminosity well above this limit {\em and} a bright $250\,\mu$m 
detection, $S_{\rm 250}=42\,$mJy. However, \citet{Page:12} surveyed a small 
area, covering just $0.12\deg^2$, and hence only probed a small volume where
galaxies as massive as the Spiderweb are rare over this volume
\citep[e.g.,][]{Fontana:06}.
We can estimate an upper limit to the duty cycle of a simultaneous
AGN and starburst activity above these limits by dividing the look-back 
time probed across $1<z<3$, $3.64\,$Gyr, by the number of X-ray AGN in 
this study, 21, and obtain a value of $170\,$Myr.
If the Spiderweb Galaxy is in a transition from merger-induced starburst 
phase to an AGN phase, and this transition is relatively short, $<170\,$Myr,
as the analysis here suggests, then the \citet{Page:12} results are 
consistent with our observations. We point out that such comparison is 
only valid if there are similar evolutionary processes going on in 
the less massive \citet{Page:12} sample.

\subsection{Proto-Cluster Environment}
\label{dis:env}

The environment of the Spiderweb Galaxy has been extensively studied
and a full review of work in this area is beyond the scope of this 
paper. We have noted previously that this galaxy hosts one of the
most powerful AGN known: both from its bolometric luminosity (indicative of
radiative power) and its radio luminosity (indicative of kinetic power).
\citet{Tanaka:10} compared massive galaxies in the environment of 
the Spiderweb Galaxy with identically selected field galaxies from the
GOODS-S field. They determined that the galaxies in the proto-cluster
environment had formed earlier and had lower SFRs, particularly close
to the HzRG and along the east-west distribution of proto-cluster members.
The radio emission has the same projected east-west axes; hence, it is possible
that the lower SFRs observed could be due to the influence of the AGN jet. 
Alternatively, they may simply be more heavily obscured.

As the Spiderweb Galaxy is situated in a well studied proto-cluster we can 
estimate its
future growth. Several authors have identified proto-cluster members and
estimated that they would merge with, or be tidally stripped by, the radio
galaxy before the present epoch. 
\citet{Hatch:09} identified 18 proto-cluster members through 
{\it HST} imaging within 150\,kpc of the central radio galaxy, the 
combined mass of which they estimated to be  
$1.9^{+1.5}_{-0.8}\times10^{11}\,M_\odot$.
\citet{Doherty:10} spectroscopically confirmed two red galaxies
within $\sim500\,$kpc\footnote{Not the $\sim300\,$kpc previously reported.}, 
in addition to those mentioned above, with stellar 
masses of the order $3-5\times10^{11}\,M_\odot$ each. 
Hence, the 
total stellar mass increase of the Spiderweb galaxy over the upcoming 
$\sim10\,$Gyr would be $\sim10^{12}\,M_\odot$, i.e., at least a doubling in 
its mass, which would put it closer to the local `$M-\sigma$' relation.
\citet{Kuiper:11} interpret their kinematic study of the environment, which
showed a double peaked velocity distribution, as being due the merger of 
two massive clusters. If each cluster contained a dominant central galaxy 
then the high AGN and starburst luminosities could well be due to a gas 
rich merger of two large galaxies. 

\section{Conclusions}
\label{conclusion}
We have measured the total IR rest-frame SED of the Spiderweb Galaxy, 
for the first time, and fitted an AGN and starburst model components.
We find that the bolometric luminosity of the AGN 
component of this source implies that the AGN has an accretion rate high 
enough to power both the jets and the observed outflow. The Spiderweb 
Galaxy  has a high SFR; however, we estimate this SFR cannot be maintained 
for long and cannot be responsible for the outflow.
We estimate the radio jets are $5-25\,$Myr old. At the current SFR, the
host galaxy has only had enough time to accumulate $\sim6\,\%$ of its 
observed stellar mass while the jets were on. Hence, the jets were likely 
triggered after most of the current burst of star formation commenced. 
The central black hole is possibly growing at a relatively faster rate than 
the host galaxy (assuming this system will end up on the local $M-\sigma$ 
relation). 
Studies of the proto-cluster environment suggest that mergers of satellites, 
and hence increased star formation, will occur on timescales of hundreds of Myr
and ultimately the stellar mass will double by the present epoch. 
We also see very tentative evidence for strong [CII]$158\mu$m emission.

\acknowledgments
NS is the recipient of an Australian Research Council Future Fellowship. 
TRG acknowledges support from the Science and Technologies Facilities Council,
as well as IDA and DARK. We thank J. Mullaney for help using the 
{\tt DecompIR} code.
This work is based in part on observations made with the {\it Spitzer Space 
Telescope}, which is operated by the Jet Propulsion Laboratory, California 
Institute of Technology under a contract with NASA. Support for this work 
was provided by NASA through an award issued by JPL/Caltech. 
HIPE is a joint development by the 
{\it Herschel} Science Ground Segment Consortium, consisting of ESA, the NASA 
{\it Herschel} Science Center, and the HIFI, PACS and SPIRE consortia

{\it Facilities:} \facility{\it Spitzer}, \facility{\it Herschel}, 
\facility{APEX}.

\clearpage
\begin{deluxetable}{lll}
\tablecolumns{3}
\tablecaption{Photometric data }
\tablewidth{0pc}
\tablehead{
\colhead{wavelength} & 
\colhead{flux density} & 
\colhead{reference}}
\startdata
$\mu$m   & mJy  & \\
\hline\
  24  &3.89$\pm$0.25& (1)\\
  100 &28.5$\pm$3.1 & (2)\\
  160 &43.0$\pm$2.9 & (2)\\
  250 &42.0$\pm$7.2 & (2)\\
  350 &38.9$\pm$7.9 & (2)\\
  500 &35.3$\pm$8.3 & (2)\\
  870 &6.7$\pm$1.3  & (3)\\
\enddata
\label{tab:data}
\tablerefs{(1) \citet{DeBreuck:10}; (2) this work ; (3) Dannerbauer et al. (2012, in preparation).}
\end{deluxetable}

\begin{deluxetable}{lll}
\tablecolumns{3}
\tablecaption{Derived physical properties}
\tablewidth{0pc}
\tablehead{
\colhead{property} & 
\colhead{value} & 
\colhead{reference}}
\startdata
  $L_{\rm IR}${\rm (Total)} & $1.97\pm0.28\times10^{13}\,L_\odot$ & (1)\\
  $L_{\rm IR}${\rm (AGN)}   & $1.17\pm0.27\times10^{13}\,L_\odot$ & (1)\\
  $L_{\rm IR}${\rm (SB)}    & $0.79\pm0.09\times10^{13}\,L_\odot$ & (1)\\
  \tableline
  $M_{\rm gal}$ (stellar)   & $\lesssim1.1\times10^{12}\,$M$_\odot$ & (2) \\
  $M_{\rm BH}$              & $\sim1\times10^{10}\,$M$_\odot$  & (3) \\
  SFR (optical)         & $57\pm8\,$M$_\odot$yr$^{-1}$  & (2)\\
  Inclination, $\theta$ & $\le68\deg$ & (4)\\
  Angular Size          & $15.8\,$arcsec & (5)\\
  $L_{\rm 2-10keV}$ (unob.) & $<1.0\times10^{12}\,L_\odot$ & (6) \\ 
  \tableline
  $L_{\rm BOL}${\rm (AGN)}  & $\sim7.0\times10^{13}\,L_\odot$ & (1)\\
  $R_{\rm Edd}$             & $\sim0.2$  & (1) \\
  SFR                   & $1390\pm150\,$M$_\odot$yr$^{-1}$  & (1)\\
  sSFR                  & $\gtrsim1.3\,$Gyr$^{-1}$  & (1) \\
  $t_{\rm char}$            & $\lesssim0.8\,$Gyr & (1)\\
  $D/sin(\theta)$       & $\sim140\,$kpc & (1)\\
  $L_{\rm 151MHz}$          & $5.6\times10^{29}\,$WHz$^{-1}$ & (1) \\
  $L_{\rm 1keV}$ (lobe)     & $1.8-4.5\times10^{11}\,L_\odot$ & (1)\\
  $t_{\rm jet}$             & $5-25\,$Myr & (1) \\
\enddata
\label{tab:res}
\tablerefs{(1) this work; (2) \citet{Hatch:09} --- note the SFR is integrated
over a 32\,kpc radius; (3) \citet{Nesvadba:11}; (4) Drouart et al. (2012, 
in preparation); (5) \citet{Carilli:97}; (6) \citet{Carilli:02a}.}
\tablecomments{Results of SED fitting are listed first, parameters from the 
literature are listed second, and derived measurements in this work are 
listed last. Total IR luminosities are integrated over $8-1000\,\mu$m.
SFR is based on the Kennicutt (1998) conversion which uses a Salpeter IMF.}
\end{deluxetable}


\clearpage

\end{document}